\documentclass[11pt,oneside]{article}
\usepackage{amsfonts,amssymb,graphicx}
\usepackage{color}
\newcommand{\Av}{{\rm Av}}

\def\be{\begin{equation}}
\def\ee{\end{equation}}
\def\bea{\begin{eqnarray}}
\def\eea{\end{eqnarray}}

\def\<{\langle}
\def\>{\rangle}
\def\~{\tilde}
\def\s{\sigma}

\def\L{\Lambda}
\def\Lp{\Lambda^\prime}

\def\b{\beta}

\def\t{\tau}

\newcommand{\qed}{\hfill \ensuremath{\Box}}
\newcommand{\R}{\Bbb R}

\newcommand{\Z}{\Bbb Z}

\newcommand{\av}[1]{\mbox{{\rm Av}}\left(#1\right)}

\newtheorem{proposition}{Proposition}
\newtheorem{theorem}{Theorem}

\newtheorem{lemma}{Lemma}

\newcommand{\proof}{{\bf Proof.}\par\noindent}
\newcommand{\beq}{\begin{eqnarray}}
\newcommand{\eeq}{\end{eqnarray}}
\newcommand{\Avk}{{\rm Av}_{\le k}}
\newcommand{\Avkp}{{\rm Av}_{k+1}}

\newcommand{\var}{\mbox{Var}}

\newcommand{\hl}{H_\Lambda}
\newcommand{\hzl}{H^0_\Lambda}
\newcommand{\hzlp}{H^0_{\Lambda^\prime}}
\newcommand{\hztlp}{\widetilde{H}^0_{\Lambda^\prime}}

\newcommand{\hlp}{H_{{\Lambda}^\prime}}
\newcommand{\hcl}{H_{{\Lambda \setminus {\Lambda}^\prime}}}

\newcommand{\hlpt}{\widetilde{H}_{{\Lambda}^\prime}}

\newcommand{\si}{\ensuremath{\sigma}}

\newcommand{\sint}{\sin t\ }
\newcommand{\sins}{\sin s\ }
\newcommand{\cost}{\cos t\ }
\newcommand{\coss}{\cos s\ }

\newcommand{\cp}{{\cal P}}
\newcommand{\cpb}{{\cal P}_\b}

\newcommand{\czb}{{\cal Z}_\beta}
\newcommand{\cx}{{\cal X}}
\newcommand{\cxz}{{\cal X}_0}
\newcommand{\cpp}{{\cal P}^{(+)}}
\newcommand{\cpm}{{\cal P}^{(-)}}
\newcommand {\rav}{{\rm Av}}
\newcommand{\jzx}{J^0_X}
\newcommand{\jx}{J_X}
\newcommand{\bla}{{\cal B}_\Lambda}
\newcommand{\blap}{{\cal B}_{\Lambda^\prime}}
\newcommand{\la}{\Lambda}
\newcommand{\lap}{\Lambda^{\prime}}
\newcommand{\vark}{{\rm V}_k}
\newcommand{\quets}[1]{\langle #1 \rangle_{t,s}}
\newcommand{\quet}[1]{\langle #1 \rangle_{t}}
\newcommand{\Cudlp}{{C}^{\Lp}_{1,2}}
\newcommand{\Cdtlp}{{C}^{\Lp}_{2,3}}
\newcommand{\Ctqlp}{{C}^{\Lp}_{3,4}}
\newcommand{\cudlp}{{c}^{\Lp}_{1,2}}
\newcommand{\cdtlp}{{c}^{\Lp}_{2,3}}
\newcommand{\ctqlp}{{c}^{\Lp}_{3,4}}
\newcommand{\Blp}{{\cal B}_{\Lambda^{\prime}}}
\begin{document}
\begin{center}
\vspace{1truecm}
{\bf\sc\Large  Interaction Flip Identities for non Centered Spin Glasses}\\
\vspace{1cm}
{Pierluigi Contucci$^{\dagger}$, Cristian Giardin\`a$^{\ddagger}$, Claudio Giberti$^{\star}$}\\
\vspace{.5cm}
{\small $\dagger$ Dipartimento di Matematica} \\
{\small Universit\`a di Bologna,40127 Bologna, Italy}\\
{\small {e-mail: {\em contucci@dm.unibo.it}}}\\
\vspace{.5cm}
{\small $\ddagger$ Dipartimento di Fisica, Informatica e Matematica}\\
{\small Universit\`a di Modena e Reggio Emilia, 41125 Modena, Italy}\\
{\small {e-mail: {\em cristian.giardina@unimore.it}}}\\
\vspace{.5cm}
{\small $\star$ Dipartimento di Scienze e Metodi dell'Ingegneria } \\
{\small Universit\`a di Modena e Reggio Emilia, 42100 Reggio Emilia, Italy}\\
{\small {e-mail: {\em claudio.giberti@unimore.it}}}\\

\vskip 1truecm
\end{center}
\vskip 1truecm
\begin{abstract}\noindent
We consider spin glass models with non-centered interactions and investigate the effect, 
on the random free energies, of flipping the interaction in a subregion of the entire volume. 
A fluctuation bound obtained by martingale methods produces, with the help of integration
by parts technique, a family of polynomial identities involving overlaps and magnetizations. 
\end{abstract}
\newpage\noindent
\section{Introduction and main results}
The study of factorisation laws for spin glass models has proved
to be a fruitful approach to investigate their low temperature phase properties.
The introduction of the concept of stochastic stability \cite{AC} and the parallel method
of the Ghirlanda-Guerra identities \cite{GG} have in fact received a lot of attention both from
the theoretical physics perspective as well as from the purely probabilistic one; 
{\color{black} in particular it is now well established that stability and factorisation properties 
are equivalent \cite{PA1,CGG1}.}
The use of those concepts has led to the final rigorous proof of the Parisi picture
for the mean field spin glass, i.e. for the Sherrington-Kirckpatrick model, by the 
work of Panchenko on ultrametricity \cite{PA2} who heavily relies on factorisation properties.

In the work \cite{CGG2} it was observed that the flip of an interaction in the zero
average spin glass case produces, by a suitable use of the integration by parts
technique, a set of identities in the deformed state. On the other hand the study of 
spin glasses is, for physical reasons, interesting also for non-zero values of the 
interaction average (\cite{deAT,N}). 
In this paper we extend the method introduced in \cite{CGG2} to the non zero average
case. In order to do so we use the technique of interpolation with the trigonometric 
method but we also show that the results remain essentially the same also for the 
linear interpolation method.

We investigate two types of flipping operations. The first flips the entire random interaction
while the second only its central part, in both cases in a subregion of the total volume. 
By comparing the random free energies of a system with given interaction to the one 
where the interactions have been flipped we show, using a martingale bound on the 
variance of their difference, that integration by parts produces a family of identities 
generalizing those found in the zero-mean interaction case.
Their first appearance comes from the physical literature where, within the 
formalism of replica quantum field theory \cite{DeDG, Te},
they are refereed as {\it replicon} type identities, 
while the standard stochastic stability argument identifies the {\it longitudinal}
type identities.
As an application, 
for the spin glass models on a regular lattice in $d$ dimension with periodic boundary conditions, 
it is interesting to consider a flip of the centered part of the interactions belonging to the boundary hyperplane 
perpendicular to one of the $d$ directions. In this case the bound on the variance of the free energies difference 
implies a bound on the {\em generalized stiffness exponents} \cite{FH}. 

The infinite volume identity obtained by flipping the entire random interaction on a subregion $\Lp$ of the lattice $\L$ reads:
\beq
& &  \b^2 \int_{0}^\pi \int_{0}^\pi  dt\, ds\, h_1(t,s) [\<m^{\Lp}_1 m^{\Lp}_2 \>_{t,s}-\<m^{\Lp}_1\>_t\< m^{\Lp}_2 \>_s]\nonumber\\
&+&2\b^3\int_{0}^\pi  \int_{0}^\pi  dt\,ds\,  h_2(t,s) \left [ \<m^{\Lp}_1 c^{\Lp}_{1,2} \>_{t,s} -  \<m^{\Lp}_1 c^{\Lp}_{2,3} \>_{t,s,t}\right ]\nonumber\\
&-& \b^4\int_{0}^\pi  \int_{0}^\pi  dt\,ds\,  k_2(t,s)\left [\<{c^{\Lp}_{1,2}}^2 \>_{t,s} -2  \<c^{\Lp}_{1,2}c^{\Lp}_{2,3} \>_{s,t,s}+
\<c^{\Lp}_{1,2}c^{\Lp}_{3,4} \>_{t,s,s,t}\right] \to0,\nonumber
\eeq
provided that in the limit the volume of $\Lp$  is non vanishing with respect to that of $\L$. In the previous equations 
$m_k^{\Lp}$ and $c_{l,k}^{\Lp}$ represent normalized magnetization and covariances inside $\Lp$ for a set of replicas 
$l,k\,...$ of the system, and the brackets denote the equilibrium states deformed according to the 
interpolating scheme which determines also the kernel functions $h_1(t,s), h_2(t,s), k_2(t,s)$. 
Similarly to the case of  the Ghirlanda-Guerra identities studied in \cite{CGN}, here the effect 
in the identities derived from spin flip due to the presence of a non-centered disorder 
is the appearance of terms depending on the system magnetization.
The simpler identity (i.e. without magnetization)  can be 
obtained, either by flipping only the centerd part of the disorder, or by taking the integral  with respect to a parameter 
tuning the averages of the disorder. In both cases only the last term of the previous identity survives,
the replicon part.

The paper is organised as follows: in the next Section we define the general class of model for which our result apply and 
we set the notations.  In Section 3, after introducing  the flipping operations and the corresponding trigonometric interpolations, 
we give the explicit expressions for the variances of the difference of pressures.  In Section 4
the self-averaging theorem is presented. It states that the flip in the sub-volume $\Lp$ of the centered part of the disorder produces an effect on 
the variance which is of the same order of the volume of $\Lp$. On the other hand, the flip of the complete disorder has a non local 
effect since in this case the variance is bounded by the whole space volume $\L$. In the final Section 5 we deduce the identities as consequences 
of the results of the previous sections.  In the Appendix we finally show that making use of a different (e.g. linear) interpolation scheme other identities
can be obtained, however  the ``core'' part of the identities involving  $\<{c_{1,2}^{\Lp}}^2\>_{t,s} -2\<c_{1,2}^{\Lp}c_{2,3}^{\Lp}\>_{t,s,t} + \<c_{1,2}^{\Lp}c_{3,4}^{\Lp}\>_{t,s,s,t}$
is still present.

\section{Definitions}
Let introduce the main quantities and the class of models that we will consider.
\begin{itemize}
\item {\it Hamiltonian}.\\ For every $\Lambda\subset \Z^d$ let
$\{H_\Lambda(\sigma)\}_{\s\in\Sigma_N}$
be a family of
$2^{|\Lambda|}$ {\em translation invariant (in distribution)
Gaussian} random variables defined, in analogy with \cite{RU}, according to
the  general representation
\be\label{ham}
H_{\Lambda}(\s) \; = \; - \sum_{X\subset \Lambda} J_X\s_X
\label{hami}
\ee
where
$$
\s_X=\prod_{i\, \in X}\s_i \; ,
$$
($\s_\emptyset=0$) and the $J$'s are independent Gaussian variables with
mean
\be\label{mean_disorder}
{\rm Av}(J_X) = \mu_X \; ,
\ee
and variance
\be\label{var_disorder}
{\rm Av}((J_X-\mu_X)^2) = \Delta^2_X  \; .
\ee
Given any subset $\Lp\subseteq \L$, we also write
\be\label{hsomma}
\hl(\s)=\hlp(\s)+\hcl(\s)
\ee
where
\be\label{hamiloc}
\hlp(\s)= - \sum_{X\subset \Lp} J_X\s_X,\quad \hcl(\s)= - \sum_{X\subset \L \atop X \nsubseteq \Lp} J_X\s_X\;.
\ee
\item {\it Average and Covariance matrix}.\\
The Hamiltonian $H_{\Lambda}(\s)$ has average
\be
{\cal B}_\Lambda(\sigma):=\av{H_\Lambda(\sigma)}=-\sum_{X\subset \Lambda} \mu_X \sigma_X
\ee
and covariance matrix
\begin{eqnarray}\label{cov-matr}
\label{cc}
{\cal C}_\Lambda (\s,\tau) \; &:= &\;
\av{(H_\Lambda(\s)-{\cal B}_\Lambda(\sigma))(H_\Lambda (\tau)-{\cal B}_\Lambda(\tau))}
\nonumber\\
& = & \; \sum_{X\subset\Lambda}\Delta^2_X\s_X\t_X\, .
\end{eqnarray}
\item{\it Thermodynamic Stability}\\
The Hamiltonian is thermodynamically stable if there exists a constant $\bar{c}$ such that
\be\label{thermstab}
\sup_{\L \subset \Z^d} \frac{1}{|\Lambda|} \sum_{X\subset \L} |\mu_X| \le \bar{c} < \infty,\quad \sup_{\L \subset \Z^d} \frac{1}{|\Lambda|} \sum_{X\subset \L} \Delta_X^2 \le \bar{c} < \infty.
\ee
Together with translation invariance a condition like (\ref{thermstab}) is equivalent to 
$$
\sum_{X \ni 0} \frac{\Delta_X^2}{|X|}\le \bar{c}
$$
or to
\be\label{thermstab2}
\sum_{\hat{X}} \Delta_{\hat{X}}^2 \le \bar{c}
\ee
where the sum is over the equivalence classese $\hat X$ of the traslation group.
Thanks to the previous relations  a thermodynamically stable model fullfills the bound
\be\label{thermstab}
{\cal B}_\Lambda(\s) \le \bar{c} |\L|,\quad {\cal C}_\Lambda (\s,\tau) \le \bar{c} |\L|
\ee
and has an order 1 normalized mean and covariance 
\be\label{bnorm}
b_{\L}(\s):=\frac{1}{|\L|} {\cal B}_\Lambda(\s) 
\ee
\be\label{cnorm}
c_\L(\s,\t):=\frac{1}{|\L|}{\cal C}_\Lambda (\s,\tau).
\ee
\item {\it Average and Covariance matrix in $\Lp$}
\be
D_{\Lp}:=\sum_{X\subset \Lp}\Delta_X^2,\quad {\cal B}_{\Lp}(\sigma):=-\sum_{X\subset \Lp} \mu_X \sigma_X,\quad
C_{\Lp}(\sigma,\tau):= \sum_{X\subset\Lp}\Delta^2_X\s_X\t_X
\ee
\be
{\cal B}_{\L\setminus \Lp}(\sigma):=-\sum_{X \nsubseteq  \Lp} \mu_X \sigma_X,\quad
C_{\L\setminus \Lp}(\sigma,\tau):= \sum_{X\nsubseteq \Lp}\Delta^2_X\s_X\t_X
\ee
\be\label{bcintens}
d_{\Lp}:=\frac{D_{\Lp}}{|\Lp|},\quad b_{\Lp}(\s):=\frac{1}{|\Lp|} {\cal B}_{\Lp}(\s),\quad  c_{\Lp}(\s,\t):=\frac{1}{|\Lp|}{\cal C}_{\Lp} (\s,\tau).
\ee
\item{\it Parametrized Hamiltonian}\\
In the following we will consider families of random  Hamiltonians $\{X_{\Lambda,t}(\sigma)\}_{\s\in\Sigma_N}$ depending on an additional parameter $t\in I$, where $I\subset \R$ is an interval. 
For a given inverse temperature $\b$, the corresponding parametrized partition function, pressure, random and quenched measure are defined:
\be\label{interpz}
\czb(t)=\sum_\sigma e^{-\b X_{\Lambda,t}(\s)},
\ee
\be\label{interpp}
\cpb(t)=\ln \czb(t),
\ee
\be\label{interpomega}
\omega_t(-)=\czb(t)^{-1}\sum_\sigma (-) e^{-\b X_{\Lambda,t}(\s)},
\ee
\be\label{interpquench}
\<-\>_t=\av{\omega_t(-)},
\ee
where $\av{\cdot}$ is the average with respect to the randomness in $X_{\Lambda,t}$.\\
The measures on the replicated system are defined as usual. For istance,  $\omega_{t,s}:=\omega_t \otimes \omega_s$ is the random interpolated state for
a two-copies system  and  $\<-\>_{t,s}=\av{\omega_{t,s}}$ is the corresponding quenched state (the dependence of the quenched states on $\b$ will be omitted).  Moreover, we will use the symbol 
$\quets{\Cudlp}$ to denote the quenched average of the covariance matrix $C_{\Lp}(\sigma,\tau)$ of two copies of the system labelled by 1 and 2, i.e.:
$$
\quets{\Cudlp}:=\av{\omega_{t,s}(C_{\Lp}(\sigma,\tau))}=\av{\czb(t)^{-1}\czb(s)^{-1}\sum_{\s,\t}C_{\Lp}(\sigma,\tau)e^{-\b X_{\Lambda,t}(\s)}e^{-\b X_{\Lambda,s}(\t)}},
$$
while for the quenched average  of ${\cal B}_{\Lp}(\sigma){\cal B}_{\Lp}(\tau)$ and ${\cal B}_{\Lp}(\sigma)C_{\Lp}(\sigma,\tau)$ we will write
$$
\quets{M^{\Lp}_1M^{\Lp}_2}:=\av{\omega_{t,s}({\cal B}_{\Lp}(\sigma){\cal B}_{\Lp}(\tau))},\quad \quets{M^{\Lp}_1\Cudlp}:=\av{\omega_{t,s}({\cal B}_{\Lp}(\sigma)C_{\Lp}(\sigma,\tau))}
$$
In the same way, we can define the quenched average over three or more copies of the system  with respect to any choice of the interpolating parameters. 
 For instance, for an ordered triple of copies (1,2,3)  the queched  average of $C_{\Lp}(\sigma,\tau)C_{\Lp}(\tau,\eta)$ with interploating parameters $(t,s,t)$ is given by
$$
\langle \Cudlp\Cdtlp \rangle_{t,s,t}:=\av{\czb(t)^{-2}\czb(s)^{-1}\sum_{\s,\t,\eta} C_{\Lp}(\sigma,\tau)C_{\Lp}(\tau,\eta)e^{-\b  X_{\Lambda,t}(s)}e^{-\b X_{\Lambda, s}(\t)} e^{-\b X_{\Lambda,t} (\eta)}}.
$$

\par\noindent
\end{itemize}

\section{Spin flip polynomials}
In this section we study the effect of flipping the disorder inside a subregion $\Lambda^\prime \subset \Lambda$.  
The disorder can be flipped in two ways, the first one being obviuosly:
\be\label{flip}
{\tt F}:\left \{
\begin{array}{l}
\jx \rightarrow -\jx,\quad \mbox{for all}\quad X\subset \lap,\\
\jx \rightarrow \jx,\quad \mbox{for all}\quad X\subset \Lambda \setminus {\Lambda}^\prime.
\end{array}
\right.
\ee
The second  one can be introduced considering the centered disorder variables $\{J^0_X\}_X$:
$$
\jzx=J_X-\mu_X,
$$
 and defining the flip as
 \be\label{flip0}
{\tt F_0}:\left \{
\begin{array}{l}
\jzx \rightarrow -\jzx,\quad \mbox{for all}\quad X\subset \lap,\\
\jzx \rightarrow \jzx,\quad \mbox{for all}\quad X\subset \Lambda \setminus {\Lambda}^\prime
\end{array}
\right.
\ee
In this second case we restrcit the flip to the centered part of the disorder. The effect  of these transformations on the Hamiltonian is obviuos. 
In fact, denoting with $\hzl$ the hamiltonian with disorder $\jzx$,
we can write
$
\hl(\s)=\hzl(\s)+\bla(\s)
$
and selecting the subvolume  $\lap$, we have 
\be\label{hhcb}
\hl(\s)=\hzlp(\s)+\blap(\s)+\hcl(\s).
\ee
Thus the action of the flips on the Hamiltonian are: 
$$
{\tt F} [\hl(\s)]=-\hzlp(\s)-\blap(\s)+\hcl(\s),\quad {\tt F_0} [\hl(\s)]=-\hzlp(\s)+\blap(\s)+\hcl(\s).
$$
We are interested in the variation of the random pressure $\cal P$ of $\hl$
\be
{\cal P}=\ln \sum_\s \exp \b(-\beta\hzlp(\s) -\blap(\s)- \hcl(\s))
\ee
 when the Hamiltonian is flipped in both the ways. 
That is, denoting with ${\cal P}^{(-)}$ the pressure of ${\tt F} [\hl(\s)]$ and with ${\cal P}^{(-)}_0$ that of ${\tt F_0}[\hl(\s)]$, i.e.
$$
{\cal P}^{(-)}=\ln \sum_\s \exp\beta (\hzlp(\s) +\blap(\s) -\hcl(\s)),\quad {\cal P}^{(-)}_0=\ln \sum_\s  \exp \b( \hzlp(\s) -\blap(\s)- \hcl(\s)),
$$
 we consider
\be\label{defchi}
\cx= {\cal P} -  {\cal P}^{(-)},\quad \cxz= {\cal P} -   {\cal P}_0^{(-)}.
\ee
In particular, we want to give an explicit expression for the variances of $\cx$ and $\cxz$. This can be obtained by introducing an interpolation path, parametrized by $t\in (a,b)$, connecting $\hl(\s)$
with ${\tt F} [\hl(\s)]$ or ${\tt F_0}[\hl(\s)]$. 
%
Let us now specify the details of the  trigonometric interpolations  that we will use. Following the approach of \cite{CGG2}, we introduce a second family  of Gaussian variables $\widetilde{J}=\{\widetilde{J}_X\}_{X\subset \Lambda}$ with
the same distribution of $J$ and independent of it:
$$
{\rm Av}(\widetilde{J}_X) = \mu_X \; ,{\rm Av}((\widetilde{J}_X-\mu_X)^2) = \Delta^2_X  \;  , \rav((J_X-\mu_X)(\widetilde{J}_Y-\mu_Y))=0
$$
and the corresponding hamiltonian
\be
\widetilde{H}_{\Lambda}(\s) \; = \; - \sum_{X\subset \Lambda} \widetilde{J}_X\s_X = \widetilde{H}_{\Lp}^0(\s)+\blap(\s)+\widetilde{H}_{\L\setminus \Lp}(\s)
\label{hamitilde}
\ee
where $\widetilde{H}_{\Lp}^0(\s)$ is, as in (\ref{hhcb}),  the Hamiltonians corresponding to $\widetilde{J}^0_X:=\widetilde{J}_X-\mu_X$.\\
The interpolation scheme that we will use depends on the flip type that we want to implement.  
\begin{itemize}
\item{\tt Flip F} In this case we consider the parametrized Hamiltonian defined, for $t\in [0,\pi]$, by
\bea\label{xf}
X_{\Lambda,t}(\s)&=&\cos t\, \hlp(\s)+\sin t\, \hlpt(\s)+\hcl(\s)\\
&=&Y^0_{\Lambda,t}(\s)+g(t)\blap(\s)+\hcl(\s)\nonumber
\eea
where
$$
Y^0_{\Lambda,t}(\s):=\cos t\, \hzlp(\s)+\sin t\, \hztlp(\s)
$$
and $g(t):=\cos t +\sin t$. Denoting with ${\cal P}(t)$  the pressure corresponding to $X_{\Lambda,t}(\s)$, we have
\be
\cx={\cal P}(0)-{\cal P}(\pi)=\int_\pi^0 \frac{d {\cal P}}{d t} dt =  \b \int_\pi^0  \left ( \sint \omega_t(\hlp)-\cost \omega_t(\hlpt) \right ) dt
\ee
where $ \omega_t(-)$ is the interpolated random state (\ref{interpomega}) with hamiltonian (\ref{xf}).
\item{\tt Flip $\tt F_0$} In this case we consider the parametrized Hamiltonian defined, for $t\in [0,\pi]$, by
\be\label{xf0}
X^0_{\Lambda,t}(\s)=Y^0_{\Lambda,t}(\s)+\blap(\s)+\hcl(\s)
\ee
with pressure ${\cal P}_0(t)$. Thus
\be
\cxz={\cal P}_0(0)-{\cal P}_0(\pi) = \b \int_\pi^0  \left ( \sint \omega_t(\hzlp)-\cost \omega_t(\hztlp) \right ) dt
\ee
where $ \omega_t(-)$ is the interpolated random state (\ref{interpomega}) with hamiltonian (\ref{xf0}).
\end{itemize}
\begin{lemma}
\label{lemma1}
For the random variable $\cxz$ we have:
\be\label{avchizero}
\av{\cxz}=0
\ee
and
\bea\label{varchi0}
\av{\cxz^2}&=& \b^2\int_0^\pi \int_0^\pi dt\, ds\, k_1(s,t)  \quets{{\Cudlp}} \nonumber \\
&-& \b^4\int_0^\pi \int_0^\pi dt\, ds\; k_2(s,t) \left ( \quets{{\Cudlp}^2} - 2\langle \Cudlp\Cdtlp \rangle_{t,s,t} + \langle \Cudlp\Ctqlp \rangle_{t,s,s,t}\right )
\eea
where
\be  
k_1(s,t)=\cos (t-s),\quad k_2(s,t)=\sin^2(t-s)
\ee 
and the interpolated quenched state in (\ref{varchi0}) corresponds to the Hamiltonian (\ref{xf0}).
\end{lemma}
\proof
Defining $\cx_0(a,b)=\cp_0(b)-\cp_0(a)$ we have immediately
\be\label{avchiex}
\av{\cx_0(a,b)}= \b \int_a^b dt \left (\sint \langle \hzlp \rangle _t -\cost \langle \hztlp\rangle_t \right )
\ee
With a simple computation involving the integration by parts for Gaussian random variables $x_1,x_2,\ldots,x_n$ with mean $\av{x_i}$ and covriances 
$\av{(x_i-\av{x_i})(x_j-\av{x_j}) }$:
\bea\label{intbyparts}
\av{x_i \psi(x_1,\ldots,x_n)}&=&\av{x_i}\av{ \psi(x_1,\ldots,x_n)}\\
&+&\sum_{j=1}^n \av{(x_i-\av{x_i})(x_j-\av{x_j}) } \av{\frac{\partial \psi(x_1,\ldots,x_n) }{\partial x_j}}\nonumber
\eea
 we obtain the quenched averages of the hamiltonians:
$$
\quet{\hzlp}=-\b  D_{\Lp} \cost  +\b \cost \langle \Cudlp \rangle_{t,t},\quad \quet{\hztlp}=-\b  D_{\Lp} \sint  +\b \sint \langle \Cudlp \rangle_{t,t}
$$
which, substituted in (\ref{avchiex}), give the result (\ref{avchizero}).\par\noindent
The computation of the variances stems from the following formula:
\bea\label{eolo}
\av{\cx_0(a,b)^2}&=&\b^2 \int_a^b \int_a^b dt\; ds \left [\phantom{\hztlp} \hspace{-0.7cm}\sint \sins  \av{\phantom{\hztlp} \hspace{-0.7cm}\omega_t  (\hzlp)\omega_s(\hzlp)} \right.\\
&-&\sint\coss    \av{\omega_t(\hzlp)\omega_s(\hztlp)} \\
&-&\cost \sins    \av{\omega_t(\hztlp)\omega_s(\hzlp)} \\
&+&\left. \cost \coss    \av{\omega_t(\hztlp)\omega_s(\hztlp)} \right ]\label{mammolo}.
\eea
The result is obtained by the explicit computation of the averages, which involve a double application of the integration by part formula (\ref{intbyparts}). The computation is long but not difficoult; we sketch it 
for the first term:
\bea
\av{\phantom{\hztlp} \hspace{-0.7cm}\omega_t  (\hzlp(\s))\omega_s(\hzlp(\t))} &=& \quets{\hlp(\s)\hlp(\t))}\nonumber\\ 
&-& \quets{\hlp(\s) \Blp(\t)}- \quets{\Blp(\s)\hlp(\t)} 
+\quets{\Blp(\s)\Blp(\t)},\nonumber
\eea
where
$$
\quets{\hlp(\s)\hlp(\t)}\equiv \av{\phantom{\hlp} \hspace{-0.7cm}\omega_t  (\hlp(\s))\omega_s(\hlp(\t))} =\sum_{\s,\t}\sum_{X,Y\subset \Lp}\s_X \t_Y\av{J_X J_Y B(\s,\t; t,s)}.
$$
In the previous line, for the sake of notation, we have introduced the symbol:
$$
B(\s,\t; t,s)=\frac{e^{-\b X^0_{\Lambda,t}(\s)} e^{-\b X^0_{\Lambda,s}(\t)}}{\czb(t)\czb(s)}.
$$
Applying (\ref{intbyparts}) twice (dropping the arguments of $B$ and denoting with $\delta_{X,Y}$ the Kronecker symbol), we obtain
\bea
\av{J_X J_Y B(\s,\t; t,s)}&=&(\mu_X\mu_Y + \delta_{X,Y}\Delta_X^2 )  \av{B} + (\mu_X\Delta_Y^2 + \mu_Y \Delta_X^2)\av{\frac{\partial B}{\partial J_Y}}\nonumber\\
&+&\Delta_X^2\Delta_Y^2\av{\frac{\partial^2 B}{\partial J_X \partial J_Y}},\nonumber
\eea
where the computation of derivatives of $B(\s,\t; t,s)$ is reduced to that of the Boltzman weights, e.g.:  
\be
\frac{\partial }{\partial J_X} \left (\frac{e^{-\b X^0_{\Lambda,t}(\s)}}{\czb(t)} \right) =\b \cos t \left (\frac{e^{-\b X^0_{\Lambda,t}(\s)}}{\czb(t)} \right) (\s_X - \omega_t(\eta_X)).
\ee

\qed
\begin{lemma} 
For the random variable $\cx$ we have:
\be
\av{\cx}=\b\int_{0}^\pi dt (\cos t - \sin t)\<M^{\Lp}\>_t
\ee
and
\beq\label{varchi}
& &\var (\cx)=\b^2\int_{0}^\pi   \int_{0}^\pi  dt\,ds\, k_1(t,s)\<C^{\Lp}_{1,2} \>_{t,s}\nonumber \\
 &+&\b^2\int_{0}^\pi  \int_{0}^\pi  dt\,ds\,  h_1(t,s) [\<M^{\Lp}_1 M^{\Lp}_2 \>_{t,s}-\<M^{\Lp}_1\>_t\< M^{\Lp}_2 \>_s]\nonumber\\
&+&2\b^3\int_{0}^\pi  \int_{0}^\pi  dt\,ds\,  h_2(t,s)\left [ \<M^{\Lp}_1 C^{\Lp}_{1,2} \>_{t,s} - \<M^{\Lp}_1 C^{\Lp}_{2,3} \>_{t,s,t}\right ]\nonumber\\
&-&\b^4\int_{0}^\pi \int_{0}^\pi  dt\,ds\,  k_2(t,s)\left [\<{C^{\Lp}_{1,2}}^2 \>_{t,s} -2  \<C^{\Lp}_{1,2}C^{\Lp}_{2,3} \>_{s,t,s}+
\<C^{\Lp}_{1,2}C^{\Lp}_{3,4} \>_{t,s,s,t}\right].
\eeq
where the interpolated quenched state correspond to the Hamiltonian (\ref{xf}) and
$$
\<M^{\Lp}\>_t:=\av{\omega_t({\cal B}_{\Lp}(\sigma))}.
$$
Moreover the kernels in the previous integrals are given by
\be
h_1(t,s)=(\cos t - \sin t)(\cos s - \sin s),\, h_2(t,s)= \sin(t-s)(\cos t - \sin t)
\ee
and $k_1(t,s)$ and $ k_2(t,s)$ are defined in Lemma 1.
\end{lemma}
\proof
The proof essentially repeats that of Lemma 1, only a changing (\ref{eolo}) - (\ref{mammolo}) where the hamiltonians $\hzlp$ and $\hztlp$ are sobstituted by $\hlp$ and $\hlpt$. Thus, the computation of $\av{\phantom{\hztlp} \hspace{-0.7cm}\omega_t  (\hlp(\s))\omega_s(\hlp(\t))}= \quets{\hlp(\s)\hlp(\t))}$ (and of similar terms) 
is identical to that of Lemma 1 and the result follows from the computation of the derivatives of $B(\s,\t; t,s)$.
\qed
\section{Flip of the disorder: self averaging bounds}
In this section we will provide bounds for the variance of the random variables $\cx_0=\cp-\cpm_0$ and $\cx=\cp-\cpm$.  The following proposition shows that while the flip 
of the centered part of the disorder variables results in a change in the variance of the order of the volume of the flipped region, the complete flip of the disorder induces an effect of the 
order of the total volume. 
\begin{proposition}
Suppose that the Hamiltonian (\ref{ham}) is thermodynamically stable, see (\ref{thermstab}). Then for every set $\Lp \subset \L$ there are positive functions $r_0(\b)$ and $r(\b)$ (independent of $\Lp$ ?) such that 
\be\label{selfchi0}
V(\cx_0)=\av{\cx_0^2}-\av{\cx_0}^2 \le r_0(\b) |\Lp|,
\ee 
and
\be\label{selfchi}
V(\cx)=\av{\cx^2}-\av{\cx}^2 \le r(\b) |\L|.
\ee
\end{proposition}
\proof
We start by proving the first statement, the second is obtained by a slight modification of the argument. 
In order to implement the martingale approach devised in \cite{CG1}, we enumerate the $N=2^{|\la|}$ disorder variables $\{ J_1, \ldots J_M, J_{M+1},\ldots J_N \}$ such that the first  $M=2^{|\lap|}$ elements correspond to the interactions inside  $\lap$.  We will denote with $\Avk$ the integration with respect to the first $k$ disorder variables and $\Av_k$ the integration with respect to the $k$-th variable.  Then we consider 
$$
A_k:= \Avk (\cxz) \equiv \Avk (\cp-\cpm_0)= \cpp_k-\cpm_k, 
$$
where $\cpp_k:=\Avk(\cp)$ and $\cpm_k:=\Avk(\cpm_0)$. Introducing
$$
\Psi_k:= A_k - A_{k+1}
$$
we can write
$$
\cxz-\Av(\cxz)=\sum_{k=0}^{N-1} \Psi_k
$$
(we assume that $\Av_{0}$ means that no integration is performed, while $\Av_N$ is $\Av$). Therefore the variance of $\cxz$ is
$$
V(\cxz)=\Av((\cxz-\Av(\cxz))^2)=\sum_{k=0}^{N-1} \Av(\Psi_k^2)+2 \sum_{k>k^\prime}\Av(\Psi_k\Psi_{k^\prime}).
$$
Since the sequence $\{A_k\}_k$ form a martingale, the sequence $\Psi_k$ is a martingale difference,  then the $\Psi_k$ are mutually orthogonal and
\be
V(\cxz)=\sum_{k=0}^{N-1} \Av(\Psi_k^2).
\ee
Now, consider $\cpp_k$ and $\cpm_k$ with $k\ge M$:
\be\label{intppiu}
\cpp_k=\int_{\R^k} \ln \sum_\s e^{-\b \hzlp(J^0_1,\ldots,J^0_M;\s)-\b \blap(\s)-\b\hcl(J_{M+1},\ldots,J_N;\s)}
\prod_{i=1}^M g_i^0(J^0_i) dJ^0_i\prod_{\ell=M+1}^k g_\ell(J_\ell) d J_\ell 
\ee
\be\label{intpmeno}
\cpm_k=\int_{\R^k} \ln \sum_\s e^{\b \hzlp(J^0_1,\ldots,J^0_M;\s)-\b \blap(\s)-\b\hcl(J_{M+1},\ldots,J_N;\s)}
\prod_{i=1}^M g_i^0(J^0_i) dJ^0_i\prod_{\ell=M+1}^k g_\ell(J_\ell) d J_\ell 
\ee
where, in the previous notation for the hamiltonians, the dependence over the disorder variables is made explicit and $ g_i^0(J^0_i)$ is the density of the centered Gaussian variable $J^0_i$, while $g_\ell(J_\ell)$ is the density of
$J_\ell$. Applying the transformation (\ref{flip}) to the  $J$ variables in (\ref{intppiu}) and from the symmetry of $g^0_i$ we obtain that (\ref{intppiu}) is transformed into (\ref{intpmeno}). Thus, $\cpp_k=\cpm_k$, for all $k\ge M$, i.e.
\be\label{azero}
A_k=0,\quad k\ge M
\ee
and
\be\label{sumpsiq}
V(\cxz)=\sum_{k=0}^{M-1} \Av(\Psi_k^2).
\ee
Now, using $A_{k+1}=\Av_{k+1}(A_k)$ and $\Av(-)=\Av[\Avkp(-)]$we have
\bea
\Av(\Psi^2_k)&=&\Av[\Avkp[(A_k-A_{k+1})^2]]=\Av[\Avkp[(A_k-\Avkp(A_{k}))^2]]\nonumber\\
&=&\Av[\Avkp(A_k^2)-(\Avkp(A_k))^2]=\Av[{\rm V}_{k+1} (A_k)].\nonumber
\eea
where   $\vark (-)$ is the variance with respect to $\Av_k$. In order to estimate $\vark (A_k)$ for $k \le M-1$,
being obviously zero for $k \ge M$, we use an interpolation argument. Thus, we introduce the interpolated hamiltonian on $\lap$ defined as:
\be\label{interph0}
H^{0,(t)}_{\lap}(\s)=-\sum_{\ell=1\atop \ell \neq k+1}^M J^0_\ell \s_\ell-( J^0_{k+1}\s_{k+1}) t,\quad t\in [0,1],\quad k\le M-1,
\ee
and
\bea
A_k(t)&=& \Avk \left [\ln \sum_\s \exp{(-\b H^{0,(t)}_{\lap}(\s) -\b \blap(\s)-\b\hcl(\s))} \right.\nonumber\\
&-& \left. \ln \sum_\s \exp{(\b H^{0,(t)}_{\lap}(\s) -\b \blap(\s)-\b\hcl(\s))} \right  ].\nonumber
\eea
Then 
$$
A_k=A_k(1)=A_k(0)+B_k
$$
where 
$$
 B_k=\int_0^1 \frac{d A_k(t)}{d t}dt=B^{(+)}_k+B^{(-)}_k
$$
\be\label{bik}
B^{(+)}_k=\beta \int_0^1 \Avk \omega^{(+)}_t(J^0_{k+1}\s_{k+1}) dt,\quad B^{(-)}_k=\beta \int_0^1 \Avk \omega^{(-)}_t(J^0_{k+1}\s_{k+1}) dt
\ee
and $ \omega^{(+)}_t(-)$ and  $ \omega^{(-)}_t(-)$ are the aveages with weights proportional to $v^{(+)}(\sigma)=\exp\b(-H^{0,(t)}_{\lap}(\s) -\blap(\s)-\hcl(\s))$ and  $v^{(-)}(\sigma)=\exp \b( H^{0,(t)}_{\lap}(\s) -\blap(\s)-\hcl(\s))$, respectively. Being $A_k(0)$ constant with respect to $J^0_{k+1}$ we have ${\rm V}_{k+1} (A_k) = {\rm V}_{k+1} (B_k)$, then
\be\label{psiquadb}
\Av(\Psi^2_k)=\Av[\Avkp(B_k^2)-(\Avkp(B_k))^2]=\Av[{\rm V}_{k+1}(B_{k})]
\ee
In order to estimate $\Av(\Psi^2_k)$ we will bound separately the two terms $\Av[\Avkp({B^{(+)}_k}^2)]$ and $\Av[(\Avkp(B^{(+)}_k))^2]$. Identical bounds will hold also for $B^{(-)}_k$. \\
Before computing the average $\Avkp$ of the quantities in (\ref{bik}), let us observe that they depend on $J_{k+1}$ through  the variable  $J_{k+1}^0=J_{k+1}-\mu_{k+1}$, which apperas
not only as the arguments of the averages  but also in the measures  $\omega^{(+)}_t$ and $\omega^{(+)}_t$,  i.e. $B^{(+)}_k=B^{(+)}_k(J_{k+1}^0)$, $B^{(-)}_k=B^{(-)}_k(J_{k+1}^0)$. Thus, denoting with $\Avkp^0$ the average with respect to $J_{k+1}^0$ and changing the varaible in the integral $\Avkp$ we have:
$\Avkp (B^{(+)}_k)=\Avkp (B^{(+)}_k(J_{k+1}-\mu_{k+1})) =\Avkp^0   (B^{(+)}_k(J_{k+1}^0))$. 
This remark allows us to  integrate with respect to the centered variable $J^0_{k+1}$ by appling the integration by parts formula  (\ref{intbyparts}), which is simpler in the case of the centered variables.\\
Now we have to estimate  $\Av[\Avkp(B^{(+)}_k{}^2)]$:
$$
\Av_{k+1}\left ( B ^{(+)}_k{}^2 \right)=  \Av_{k+1}\int_0^1 \int_0^1 \Avk \omega^{(+)}_t(J_{k+1}^0\s_{k+1}) \Avk \omega^{(+)}_s(J_{k+1}^0\s_{k+1}) dt\, ds.
$$
Making use of $\Av_{k+1}\left ( B ^{(+)}_k{}^2 \right )= \Av_{k+1}^0\left ( B ^{(+)}_k{}^2 \right )$,
 applying  twice the integration by parts w.r.t. $J^0_{k+1}$ and recalling that ${\rm V}_{k+1}(B^{(+)}_k)\le \Av_{k+1}\left ({B^{(+)}_k}^2\right )$  we obtain:
\bea
&&\Av_{k+1}\left ({B^{(+)}_k}^2 \right)=\b^2 \Delta_{k+1}^2 \Avkp   \int_0^1 \int_0^1  \Avk \omega^{(+)}_t(\s_{k+1}) \Avk \omega^{(+)}_s(\s_{k+1})\,t s\, dt\, ds\nonumber\\
&-& 2 \beta^4 \Delta_{k+1}^4    \Avkp \int_0^1 \int_0^1    \Avk [\omega^{(+)}_t(\s_{k+1}) (1-\omega^{(+)}_t(\s_{k+1})^2)]\Avk [\omega^{(+)}_s(\s_{k+1})] t^2 dt\, ds \nonumber \\
&-& 2 \beta^4 \Delta_{k+1}^4    \Avkp \int_0^1 \int_0^1    \Avk [\omega^{(+)}_s(\s_{k+1}) (1-\omega^{(+)}_s(\s_{k+1})^2)]\Avk [\omega^{(+)}_t(\s_{k+1})] s^2 dt\, ds \nonumber \\
&+& 2 \beta^4 \Delta_{k+1}^4    \Avkp \int_0^1 \int_0^1    \Avk [ (1-\omega^{(+)}_t(\s_{k+1})^2)]\Avk [1-\omega^{(+)}_s(\s_{k+1})] t s dt\, ds \\
\nonumber
\eea
which implies
$$
{\rm V}_{k+1}(B^{(+)}_k)\le \frac 1 4\b^2 \Delta_{k+1}^2+ \frac{11}{6} \b^4 \Delta_{k+1}^4,
$$
(the same bound holds for $B^{(-)}$).

Finally, we have that ${\rm V}_{k+1}(B_k)={\rm V}_{k+1}(B^{(+)}_k)+{\rm V}_{k+1}(B^{(-)}_k)+2\, {\rm Cov}_{k+1}(B^{(+)}_k,B^{(-)}_k)$,  where ${\rm Cov}_{k+1}$, the covariance w.r.t. $J^0_{k+1}$, can be estimated
using the  Cauchy-Schwartz inequality. Thus we obtain
\be
{\rm V}_{k+1}(B_k)\le \b^2 \Delta_{k+1}^2+ \frac{22}{3} \b^4 \Delta_{k+1}^4,
\ee
 and recalling (\ref{sumpsiq}) and (\ref{psiquadb})
\be\label{stimachi0}
V(\cx_0)\le   \b^2 \sum_{k=0}^{M-1}\Delta_{k+1}^2+ \frac{22}{3} \b^4 \sum_{k=0}^{M-1} \Delta_{k+1}^4.
\ee
The sums in the previous inequality are taken over the subsets of $\lap$, thus we can rewrite them as:
$$
V(\cx_0)\le   \b^2 \sum_{X \subset \lap }\Delta_{X}^2+ \frac{25}{3} \b^4 \sum_{X \subset \lap} \Delta_{k+1}^4.
$$
Applying the Thermodynamic  Stability condition with the formulation (\ref{thermstab2}), ane using the inequality 
$\sum_{\hat X} \Delta_{\hat X}^4\le (\sum_{\hat X} \Delta_{\hat X}^2)^2$, we obtain
$$
V(\cx_0)\le r_0(\b) |\lap| ,
$$ 
which proves the self averaging bounds of $\cx_0$, (\ref{selfchi0}) with $r_0(\b)=( \b^2  \bar{c}+ \frac{22}{3} \b^4 \bar{c}^2) $.\\
The proof of (\ref{selfchi}) runs parallel to that of (\ref{selfchi0}) up to equation (\ref{azero}), which is no loger valid in this case.  
In fact applying  (\ref{flip0})  the integral (\ref{intppiu}) is not transformed into (\ref{intpmeno}), because of the change of sign in $\blap(\s)$.  
Thus $V(\cx)$ is in general a sum of the $N=2^{|\L|}$ terms $\av{\Psi^2_k}$ that can be estimated\footnote{ In the interpolation 
(\ref{interph0}) $J_X^0$ is substituded by $J_X$.}, as in the previous case, by using the thermodynamic stability condition (\ref{thermstab}). 
As shown above, the result is the existence of a positve function $r(\b)$ such that
\be
V(\cx)\le r(\b)  |\L|,
\ee
which concludes the proof.
\qed
\section{Spin flip identities}
\begin{theorem}\label{teorema1}
Suppose that the Hamiltonian (\ref{ham}) is thermodynamically stable, see (\ref{thermstab}). Then the following facts hold:
\begin{enumerate}
\item Consider the interpolating  quenched state corresponding to the Hamiltonian (\ref{xf0}), then
\be\label{ident0}
\lim_{\L,\Lp\nearrow \Z^d}\int_0^\pi \int_0^\pi dt\, ds\; k_2(t,s)\left ( \quets{({\cudlp})^2} - 2\langle \cudlp\cdtlp \rangle_{t,s,t} + \langle \cudlp\ctqlp \rangle_{t,s,s,t}\right )=0
\ee
where $\quets{({\cudlp})^2}$ (and analogously for the other terms)  is defined, see (\ref{bcintens}), as
$$
\quets{({\cudlp})^2}=\av{\omega^{(\b)}_{t,s}(c_{\Lp}(\sigma,\tau))}
$$
while $k_2(t,s)$ is defined in Lemma 1.
\item Consider the interpolating  quenched state corresponding to the Hamiltonian (\ref{xf}),   then for any $a > 0$ we have
\beq
\label{id-general}
& & \lim_{{ \L,\Lp\nearrow \Z^d} \atop { |\Lp|/|\L|\rightarrow  a}}\left \{ \b^2\int_{0}^\pi \int_{0}^\pi  dt\, ds\, h_1(t,s) [\<m^{\Lp}_1 m^{\Lp}_2 \>_{t,s}-\<m^{\Lp}_1\>_t\< m^{\Lp}_2 \>_s]\right.\\
&+&2\b^3 \int_{0}^\pi  \int_{0}^\pi  dt\,ds\,  h_2(t,s) \left [ \<m^{\Lp}_1 c^{\Lp}_{1,2} \>_{t,s} -  \<m^{\Lp}_1 c^{\Lp}_{2,3} \>_{t,s,t}\right ]\nonumber\\
&-&\b^4\left. \int_{0}^\pi  \int_{0}^\pi  dt\,ds\,  k_2(t,s)\left [\<{c^{\Lp}_{1,2}}^2 \>_{t,s} -2  \<c^{\Lp}_{1,2}c^{\Lp}_{2,3} \>_{s,t,s}+
\<c^{\Lp}_{1,2}c^{\Lp}_{3,4} \>_{t,s,s,t}\right] \right \}=0.\nonumber
\eeq
where $\quets{m^{\Lp}_1 m^{\Lp}_2}$ (and analogously for the other terms) is defined, see (\ref{bcintens}),  as
$$
\quets{m^{\Lp}_1 m^{\Lp}_2}=\av{\omega^{(\b)}_{t,s}(b_{\Lp}(\s) b_{\Lp}(\tau))}
$$
and functions $h_1(t,s)$, $h_2(t,s)$, $k_2(t,s)$ are defined in Lemma 2.
\end{enumerate}
\end{theorem}
\proof
Writing (\ref{varchi0}) as a funcion of the normalized quantities (\ref{bnorm}) and (\ref{cnorm}) and using the self-averaging bound (\ref{selfchi0}), we have
\bea
V(\cx_0)&=& |\Lp|\b^2\int_0^\pi \int_0^\pi dt\, ds\, \cos(t-s)  \quets{{\cudlp}} \nonumber \\
&-&  |\Lp|^2\b^4\int_0^\pi \int_0^\pi dt\, ds\; \sin^2(t-s) \left ( \quets{{\cudlp}^2} - 2\langle \cudlp\cdtlp \rangle_{t,s,t} + \langle \cudlp\ctqlp \rangle_{t,s,s,t}\right )\le r_0(\beta) |\Lp|.\nonumber
\eea
Then, the first statement follows taking the limit $\Lp\nearrow \Z^d$.\\
The proof ot the second statement is similar. We conisder (\ref{varchi})  with normalized quantities and apply (\ref{selfchi}) obtaining 
\beq\label{varchi1}
\var (\cx)&=&|\Lp|\b^2\int_{0}^\pi   \int_{0}^\pi  dt\,ds\, k_1(t,s)\<c^{\Lp}_{1,2} \>_{t,s}\\
& +& |\Lp|^2 \b^2\int_{0}^\pi  \int_{0}^\pi  dt\,ds\, h_1(t,s) [\<m^{\Lp}_1 m^{\Lp}_2 \>_{t,s}-\<m^{\Lp}_1\>_t\< m^{\Lp}_2 \>_s]\nonumber\\
&+&2 |\Lp|^2\b^3\int_{0}^\pi  \int_{0}^\pi  dt\,ds\, h_2(t,s) \left [ \<m^{\Lp}_1 c^{\Lp}_{1,2} \>_{t,s} -  \<m^{\Lp}_1 c^{\Lp}_{2,3} \>_{t,s,t}\right ]\nonumber\\
&-& |\Lp|^2\b^4\int_{0}^\pi \int_{0}^\pi  dt\,ds\, k_2(t,s)\left [\<{c^{\Lp}_{1,2}}^2 \>_{t,s} -2  \<c^{\Lp}_{1,2}c^{\Lp}_{2,3} \>_{s,t,s}+
\<c^{\Lp}_{1,2}c^{\Lp}_{3,4} \>_{t,s,s,t}\right] \le r(\beta) |\L|.\nonumber 
\eeq
Dividing the two terms of the previous inequality by $|\L|^2$ and letting $\L,\, \Lp \nearrow \Z^d$ with the constraint $|\Lp|/|\L|\rightarrow a > 0$ we obtain the result.
\qed
\\
Thus, the complete disorder flip  ($\tt F$) produces an identity which is more complex than the one obtained by flipping  the centered part of the disorder only (i.e.$\tt F_0$). Indeed, while
(\ref{ident0}) seems to be independet of the filp type and, to some extent, also on the interpolation path, the  identity  (\ref{id-general}) involves extra terms containing the 
generalised magnetizations $m$.\\ 
It is  possible to remove these terms by applying an extra average over the mean value of the disorder variables \cite{CGN} in $\Lp$. In fact, let us introduce a new parameter $\mu$ and write the averages of $J_X$ as
\be\label{parammu}
\mu_X= \mu \mu_X^\prime \quad \mbox{ for }\quad X\subset \Lp.
\ee
The effect of (\ref{parammu}) is to introduce a new parameter in  the random and interpolating quenched averages.  The latter will 
be denoted  by $\langle \cdot \rangle_{t;\mu}$ or $\langle \cdot \rangle_{t,s;\mu}$ etc... . \\
The next result states that, in $\mu$-average, the fluctuations with respect to  $\langle \cdot \rangle_{t;\mu}$ of the generalized magnetization in any macroscopic subregion $\Lp$ of $\L$  is vanishing  the large volume limit.
\begin{lemma}
For every interval $[\mu_1,\mu_2]$  and any $t$ and $\Lp\subset \L$, we have 
\be\label{varmzero}
\lim_{\Lp \nearrow \Z^d}\int_{\mu_1}^{\mu_2} d\mu \left ( \langle {m^{\Lp}}^2 \rangle_{t; \mu} - \langle m^{\Lp} {\rangle^2 _{t;\mu}}\right )=0.
\ee
\end{lemma}
The proof of this lemma runs parallel, with the obvious modifications, to that  of Lemma 4.8 of \cite{CGN} where the fluctuations of the magnetization of the whole region $\L$ and with respect the quanched state is considered. \qed\\\\
It is obvious that Lemma 2, Proposition 4.1 and thus Theorem 1 also hold  for the $\mu$-dependent random and quanched measures. 
Therefore, taking the integral of (\ref{varchi}) with respect to $\mu$,  we obtain the following:
\begin{theorem}\label{teorema2}
Suppose that the Hamiltonian (\ref{ham}) is thermodynamically stable, see (\ref{thermstab}), and consider the interpolating  quenched state corresponding to the Hamiltonian (\ref{xf}), with the averages of the disorder variables parametrizated by $\mu$, see (\ref{parammu}).
 Then for any $a > 0$ and  for any interval $[\mu_1,\mu_2]$, we have:
\be\label{identchim0}
\lim_{{ \L,\Lp\nearrow \Z^d} \atop { |\Lp|/|\L|\rightarrow  a}}  \int_{\mu_1}^{\mu_2 } d\mu \int_{0}^\pi   \int_{0}^\pi  dt\, ds\,   k_2(t,s)\left [\<{c^{\Lp}_{1,2}}^2 \>_{t,s ; \mu} -2  \<c^{\Lp}_{1,2}c^{\Lp}_{2,3} \>_{s,t,s ; \mu}+
\<c^{\Lp}_{1,2}c^{\Lp}_{3,4} \>_{t,s,s,t ; \mu} \right]  =0,
\ee
where $k_2(t,s)$ is defined in Lemma 2.
\end{theorem}
\proof
Consider the right-hand side of (\ref{varchi1}) but with the $t,s,\mu$-dependent states instead of $t,s$-dependent states, and denote with $I_1(\mu)$,
$I_2(\mu)$, $I_3(\mu)$, $I_4(\mu)$ the duoble integrals in $s,t$-variables. Integrating on $(\mu_1,\mu_2)$ we have
$$
0\le \frac{|\Lp|}{|\L|}\int_{\mu_1}^{\mu_2} d\mu\, I_1(\mu)+\frac{|\Lp|^2}{|\L|}\int_{\mu_1}^{\mu_2} d\mu\, (I_2(\mu) +2 I_3(\mu) -I_4(\mu))\le r(\beta)(\mu_2-\mu_1),
$$
which shows that
$$
\lim_{{ \L,\Lp\nearrow \Z^d} \atop { |\Lp|/|\L|\rightarrow  a}} \int_{\mu_1}^{\mu_2} d\mu\, (I_2(\mu) +2 I_3(\mu) -I_4(\mu))=0.
$$
Now we want  show that the integrals of $I_2(\mu)$, $I_3(\mu)$ vanish in the large volume  limit, thus proving (\ref{identchim0}) .  In fact, let us consider the covariance  
${\rm Cov}_{t,s ; \mu}( m^{\Lp}_1, m^{\Lp}_2 )$$:= \<m^{\Lp}_1 m^{\Lp}_2 \>_{t,s ; \mu}-\<m^{\Lp}_1\>_{t ; \mu}\< m^{\Lp}_2 \>_{s ; \mu}$ and write
\begin{eqnarray}
&&\left |  \int_{\mu_1}^{\mu_2} d\mu\, I_2(\mu)  \right | \le \int_{0}^\pi   \int_{0}^\pi  dt\, ds\, |h_1(t,s)|  \int_{\mu_1}^{\mu_2} d\mu \left | Cov_{t,s ; \mu}( m^{\Lp}_1, m^{\Lp}_2 )\right |\nonumber\\
&\le& \int_{0}^\pi   \int_{0}^\pi  dt\, ds\, |h_1(t,s)|  \sqrt{ \int_{\mu_1}^{\mu_2} d\mu\, Var_{t ; \mu} (m^{\Lp}_1)}\sqrt{ \int_{\mu_1}^{\mu_2} d\mu\, Var_{s ; \mu} (m^{\Lp}_2)}\nonumber\\
&\le& 4 \pi^2\max_{t\in [0,\pi]} \int_{\mu_1}^{\mu_2} d\mu\, Var_{t ; \mu} (m^{\Lp}_1)\rightarrow 0,\nonumber
\end{eqnarray}
where the Schwarz inequality and (\ref{varmzero}) have been used.
In order to bound the integral of $I_3(\mu)$ we introduce the function $\bar{m}(t,\mu):=\< m^{\Lp} \>_{t; \mu}$ and write
$$
 \<m^{\Lp}_1 c^{\Lp}_{1,2} \>_{t,s ; \mu} -  \<m^{\Lp}_1 c^{\Lp}_{2,3} \>_{t,s,t ; \mu } = \<(m^{\Lp}_1 - \bar{m}(t,\mu))c^{\Lp}_{1,2} \>_{t,s ; \mu} 
+ \<(\bar{m}(t,\mu)-m^{\Lp}_1) c^{\Lp}_{2,3} \>_{t,s,t ; \mu}
$$
thus
\begin{eqnarray}
&&\left |  \int_{\mu_1}^{\mu_2} d\mu\, I_3(\mu)  \right | \le \int_{0}^\pi   \int_{0}^\pi  dt\, ds\, |h_2(t,s)|  \int_{\mu_1}^{\mu_2} d\mu \left |\<m^{\Lp}_1 c^{\Lp}_{1,2} \>_{t,s ; \mu} -  \<m^{\Lp}_1 c^{\Lp}_{2,3} \>_{t,s,t ; \mu } \right | \\ \nonumber\\
&\le& 2\pi^2 \int_{\mu_1}^{\mu_2} d\mu \left |\<(m^{\Lp}_1 - \bar{m}(t,\mu)) c^{\Lp}_{1,2} \>_{t,s ; \mu} \right | + 2 \pi^2 \int_{\mu_1}^{\mu_2} d\mu \left |   \<  (\bar{m}(t,\mu)-m^{\Lp}_1)    c^{\Lp}_{2,3} \>_{t,s,t ; \mu } \right |  \nonumber \\
&\le& 4 \pi^2 \bar{c} \max_{t\in [0,\pi]} \int_{\mu_1}^{\mu_2} d\mu\, Var_{t ;\mu} (m^{\Lp}_1)\rightarrow 0\nonumber
\end{eqnarray}
where  the Schwarz inequality, (\ref{varmzero}) and boundedness of $c^{\Lp}_{i,j}$ have been used.
\qed
\\
{\color{black}
{\bf Remark.} As it was already observed in \cite{CGG2}, the terms appearing in the polynomial $\omega_{t,s}((c^{\Lp}_{1,2})^2)-2\;\omega_{s,t,s}(c^{\Lp}_{1,2}c^{\Lp}_{2,3})+\omega_{t,s,t,s}(c^{\Lp}_{1,2}c^{\Lp}_{3,4})$ when expressed 
in terms of the spin variables, read as:
\beq\label{nonnapapera}
\omega_{t,s}((c^{\Lp}_{1,2})^2)=\frac{1}{|\Lp|^2}\sum_{X,Y\subset \Lp} \Delta_X^2 \Delta_Y^2
\omega_t(\si_X^{(1)}\si_Y^{(1)})\omega_s(\si_X^{(2)}\si_Y^{(2)}),\nonumber\\
\omega_{s,t,s}(c^{\Lp}_{1,2}c^{\Lp}_{2,3})=\frac{1}{|\Lp|^2}\sum_{X,Y\subset \Lp} \Delta_X^2 \Delta_Y^2
\omega_s(\si_X^{(1)})\omega_t(\si_X^{(2)}\si_Y^{(2)})\omega_s(\si_Y^{(3)}),\nonumber\\
\omega_{t,s,s,t}(c^{\Lp}_{1,2}c^{\Lp}_{3,4})=\frac{1}{|\Lp|^2}\sum_{X,Y\subset \Lp} \Delta_X^2 \Delta_Y^2
\omega_t(\si_X^{(1)})\omega_s(\si_X^{(2)})\omega_s(\si_Y^{(3)})\omega_t(\si_Y^{(4)}),\nonumber
\eeq
thus
\beq\label{pluto}
& &\omega_{t,s}((c^{\Lp}_{1,2})^2)-2\;\omega_{s,t,s}(c^{\Lp}_{1,2}c^{\Lp}_{2,3})+\omega_{t,s,t,s}(c^{\Lp}_{1,2}c^{\Lp}_{3,4})=\\
& &\frac{1}{|\Lp|^2}\sum_{X,Y\subset \Lp} \Delta_X^2 \Delta_Y^2 \left [\omega_{t}(\si_X\si_Y)- \omega_{t}(\si_X)\omega_{t}(\si_Y)\right ]
\left [\omega_{s}(\si_X\si_Y)- \omega_{s}(\si_X)\omega_{s}(\si_Y)\right ]\nonumber
\eeq
(with the replica indices dropped).  These expressions make clear the correlation-like structure of these quantities. In the case of the 
Edwards-Anderson model \cite{CG2}, (\ref{pluto}) has a form which is similar to that of spin-glass susceptibility which, in turn,  
is related to the replicon mass \cite{Te}.\\
\vspace{.5cm}
\paragraph*{Acknowledgements.}
We acknowledge  financial support by the Italian Research Funding Agency (MIUR)
through  PRIN project ``Statistical mechanics of disordered and complex systems'', grant n. 2010HXAW77010 
and  FIRB project  ``Stochastic processes in interacting particle systems: duality, metastability and their applications'',   
grant n.\ RBFR10N90W and the Fondazione Cassa di Risparmio Modena through the International Research 2010 
project.
}

\section{Appendix: linear interpolation}
In this appendix, to implement the flip $\tt F$, we consider an interpolation scheme which is  simpler than those used in the previous sections since it does not require a second set of disorder variables. Indeed, here interpolation is linear over the flull disorder variables of the flipped region, i.e.:
\be\label{interplin}
X^l_{\Lambda,t}(\s)=t\hlp(\s)+\hcl(\s),\quad  t\in [-1,1].
\ee
{\bf Remark} This interpolation is singular in $t=0$, in the sense that for this value of the parameter there is no interaction inside $\Lp$.\\
{\color{black}The following lemma  shows that  this  straightforward 
interpolation scheme actually produces an expression  for the variance of $\cx$  which is more complex than the ones obtained with the trigonometric interpolations (\ref{varchi0}),(\ref{varchi}). In fact, introducing the pressure (\ref{interpp}) with  (\ref{interplin}),  using the integral representation for  the difference of pressure (\ref{defchi}), $\cx=\int_{-1}^{1}   \frac{d {\cal P}(t)}{d t} dt$, and applying  the integration by parts formula, we obtain:} 
\begin{lemma}
For the random variable $\cx$ and the interpolation scheme (\ref{interplin}) we have 
$$
\av{\cx}=-\b \int_{-1}^1 dt\, (\<M^{\Lp}\>_t  + \b t \<C^{\Lp}_{1,2} \>_{t,t})
$$
and
\beq
& &\var(\cx)=\b^2\int_{-1}^1   \int_{-1}^1   dt\,ds\;\<C_{1,2}^{\Lp}\>_{t,s}-2\b^4\int_{-1}^1   \int_{-1}^1   dt\,ds \;  s\,t\ \<C_{1,2}^{\Lp}\>_{t,s}\nonumber\\
&+&\b^2\int_{-1}^1 \int_{-1}^1   dt\,ds\;\left(\<M_1^{\Lp}M_2^{\Lp}\>_{t,s}-\<M^{\Lp}\>_t\<M^{\Lp}\>_s\right)\nonumber\\
&-&2\b^3 \int_{-1}^1   \int_{-1}^1  dt\,ds\; t\left(\<M_1^{\Lp}C_{1,2}^{\Lp}\>_{t,s}-\<C_{1,2}^{\Lp}\>_{t,t}\<M^{\Lp}\>_s\right)\nonumber\\
&+&2\b^3\int_{-1}^1  \int_{-1}^1   dt\,ds\; t\left(\<M_1^{\Lp}C_{2,3}^{\Lp}\>_{t,s,t}+\<M_1^{\Lp}C_{2,3}^{\Lp}\>_{s,t,t}-D_{\Lp}\<M^{\Lp}\>_{s}\right)\nonumber\\
&+&\b^4\int_{-1}^1   \int_{-1}^1  dt\,ds\; s\,t\left(\<C_{1,2}^{\Lp}C_{3,4}^{\Lp}\>_{t,t,s,s}-\<C_{1,2}^{\Lp}\>_{tt}\<C_{1,2}^{\Lp}\>_{s,s}\right)\nonumber\\
&-&4\b^4\int_{-1}^1  \int_{-1}^1   dt\,ds\; t^2\left(\<C_{1,2}^{\Lp}C_{2,3}^{\Lp}\>_{t,t,s}-\<C_{1,2}^{\Lp}C_{3,4}^{\Lp}\>_{t,t,s,t}\right)\nonumber\\
&+&\b^4\int_{-1}^1   \int_{-1}^1  dt\,ds\; ts \left(\<{C_{1,2}^{\Lp}}^2\>_{t,s} -2\<C_{1,2}^{\Lp}C_{2,3}^{\Lp}\>_{t,s,t} + \<C_{1,2}^{\Lp}C_{3,4}^{\Lp}\>_{t,s,s,t}\right).\nonumber
\eeq
where the interpolated quenched states correspond to the Hamiltonian (\ref{interplin}).
\end{lemma}
{\bf Proof:} same as in Lemma  \ref{lemma1}. 
\qed\\
Arguing as in Theorems \ref{teorema1} and \ref{teorema2},  we obtain:
\begin{theorem}
Assume that the disorder satisfies the Thermodynamic Stability property, see (\ref{thermstab}), and consider the interpolating  quenched state corresponding to the Hamiltonian (\ref{interplin}),   then for any $a > 0$ we have
\beq
& & \lim_{{ \L,\Lp\nearrow \Z^d} \atop { |\Lp|/|\L|\rightarrow  a}}\left \{ \b^2\int_{-1}^{1} \int_{-1}^1  dt\,  ds\, 
 \left(\<m_1^{\Lp}m_2^{\Lp}\>_{t,s}-\<m^{\Lp}\>_t\<m^{\Lp}\>_s\right)\right.\nonumber\\
&-&2\b^3\int_{-1}^1  \int_{-1}^1  dt\,  ds\; t\left(\<m_1^{\Lp}c_{1,2}^{\Lp}\>_{t,s}-\<c_{1,2}^{\Lp}\>_{t,t}\<m^{\Lp}\>_s\right)\nonumber\\
&+&2\b^3\int_{-1}^1   \int_{-1}^1  dt\,  ds\; t\left(\<m_1^{\Lp}c_{2,3}^{\Lp}\>_{t,s,t}+ \<m_1^{\Lp}c_{2,3}^{\Lp}\>_{s,t,t}  -d_{\Lp}  \<m^{\Lp}\>_{s}\right)\nonumber\\
&+&\b^4\int_{-1}^1  \int_{-1}^1  dt\,  ds\; ts\left(\<c_{1,2}^{\Lp}c_{3,4}^{\Lp}\>_{t,t,s,s}-\<c_{1,2}^{\Lp}\>_{t,t}\<c_{1,2}^{\Lp}\>_{s,s}\right)\nonumber\\
&-&4\b^4\int_{-1}^1  \int_{-1}^1  dt\,  ds\; t^2\left(\<c_{1,2}^{\Lp}c_{2,3}^{\Lp}\>_{t,t,s}-\<c_{1,2}^{\Lp}c_{3,4}^{\Lp}\>_{t,t,s,t}\right)\nonumber\\
&+&\b^4\left.\int_{-1}^1 \int_{-1}^1  dt\,  ds\; ts \left(\<{c_{1,2}^{\Lp}}^2\>_{t,s} -2\<c_{1,2}^{\Lp}c_{2,3}^{\Lp}\>_{t,s,t} + \<c_{1,2}^{\Lp}c_{3,4}^{\Lp}\>_{t,s,s,t}\right)\nonumber
\right\}=0.
\eeq
Moreover, in the same hypotheses, introducing the parametrization (\ref{parammu}) and the parametrized  deformed states $\langle \cdot \rangle_{t;\mu}$ or $\langle \cdot \rangle_{t,s;\mu}$ etc...
we have for any interval $[\mu_1,\mu_2]$:
\beq
&&\lim_{{ \L,\Lp\nearrow \Z^d} \atop { |\Lp|/|\L|\rightarrow  a}}\int_{\mu_1}^{\mu_2 } d\mu \left \{\b^4 \int_{-1}^1  \int_{-1}^1  dt\,  ds\; ts\left(\<c_{1,2}^{\Lp}c_{3,4}^{\Lp}\>_{t,t,s,s;\mu}-\<c_{1,2}^{\Lp}\>_{t,t;\mu}\<c_{1,2}^{\Lp}\>_{s,s;\mu}\right)\right.\nonumber\\
&-&4\b^4\int_{-1}^1  \int_{-1}^1  dt\,  ds\; t^2\left(\<c_{1,2}^{\Lp}c_{2,3}^{\Lp}\>_{t,t,s;\mu}-\<c_{1,2}^{\Lp}c_{3,4}^{\Lp}\>_{t,t,s,t;\mu}\right)\nonumber\\ 
& + &\left. \b^4 \int_{-1}^1   \int_{-1}^1  dt\, ds\,  ts \left (\<{c^{\Lp}_{1,2}}^2 \>_{t,s ; \mu} -2  \<c^{\Lp}_{1,2}c^{\Lp}_{2,3} \>_{s,t,s ; \mu}+
\<c^{\Lp}_{1,2}c^{\Lp}_{3,4} \>_{t,s,s,t ; \mu} \right) \right \} =0.\nonumber
\eeq
\end{theorem}

\end{document}